\documentstyle[12pt]{article}

\catcode`@=11
\newcount\@tempcntc
\def\@citex[#1]#2{\if@filesw\immediate\write\@auxout{\string\citation{#2}}\fi
  \@tempcnta\z@\@tempcntb\m@ne\def\@citea{}\@cite{\@for\@citeb:=#2\do
    {\@ifundefined
       {b@\@citeb}{\@citeo\@tempcntb\m@ne\@citea\def\@citea{,}{\bf ?}\@warning
       {Citation `\@citeb' on page \thepage \space undefined}}%
    {\setbox\z@\hbox{\global\@tempcntc0\csname b@\@citeb\endcsname\relax}%
     \ifnum\@tempcntc=\z@ \@citeo\@tempcntb\m@ne
       \@citea\def\@citea{,}\hbox{\csname b@\@citeb\endcsname}%
     \else
      \advance\@tempcntb\@ne
      \ifnum\@tempcntb=\@tempcntc
      \else\advance\@tempcntb\m@ne\@citeo
      \@tempcnta\@tempcntc\@tempcntb\@tempcntc\fi\fi}}\@citeo}{#1}}
\def\@citeo{\ifnum\@tempcnta>\@tempcntb\else\@citea\def\@citea{,}%
  \ifnum\@tempcnta=\@tempcntb\the\@tempcnta\else
   {\advance\@tempcnta\@ne\ifnum\@tempcnta=\@tempcntb \else \def\@citea{--}\fi
    \advance\@tempcnta\m@ne\the\@tempcnta\@citea\the\@tempcntb}\fi\fi}
\catcode`@=12

\begin{document}

\title{\vskip-3cm{\baselineskip14pt
\centerline{\normalsize DESY 02-134\hfill ISSN 0418-9833}
\centerline{\normalsize hep-ph/0210161\hfill}
\centerline{\normalsize October 2002\hfill}}
\vskip1.5cm
Heavy-quarkonium creation and annihilation with
${\cal O}(\alpha_s^3\ln\alpha_s)$ accuracy}
\author{{\sc Bernd A. Kniehl}, {\sc Alexander A. Penin}\thanks{Permanent
address: Institute for Nuclear Research, Russian Academy of Sciences,
60th October Anniversary Prospect 7a, 117312 Moscow, Russia.},
{\sc Matthias Steinhauser}\\
{\normalsize II. Institut f\"ur Theoretische Physik, Universit\"at Hamburg,}\\
{\normalsize Luruper Chaussee 149, 22761 Hamburg, Germany}\\
\\
{\sc Vladimir A. Smirnov}\\
{\normalsize Nuclear Physics Institute, Moscow State University,}\\
{\normalsize 119899 Moscow, Russia}}

\date{}

\maketitle

\thispagestyle{empty}

\begin{abstract}
We calculate the ${\cal O}(\alpha_s^3\ln\alpha_s)$ contributions to the
heavy-quarkonium production and annihilation rates. 
Our result sheds new light on the structure of the high-order perturbative
corrections and opens a new perspective for a high-precision theoretical
analysis.
We also determine the three-loop anomalous dimensions of the nonrelativistic
vector and pseudoscalar currents. 
\medskip

\end{abstract}

\newpage

The theoretical study of nonrelativistic heavy-quark-antiquark systems is
among the earliest applications of perturbative quantum chromodynamics (QCD)
\cite{AppPol} and has by now become a classical problem. 
Its applications to bottomonium \cite{NSVZ} and top-antitop \cite{FadKho}
physics entirely rely on the first principles of QCD.
These systems allow for a model-independent perturbative treatment.
Nonperturbative effects \cite{VolLeu} are well under control for the
top-antitop system and, at least within the sum-rule approach, also for
bottomonium.
This makes heavy-quark-antiquark systems an ideal laboratory to determine
fundamental parameters of QCD, such as the strong-coupling constant $\alpha_s$
and the heavy-quark masses $m_q$.
The bottom-quark mass $m_b$ is of particular interest, in view of current and
future $B$-physics experiments.
In the observables employed to extract the Cabibbo-Kobayashi-Maskawa matrix
elements and to gain deeper insight in the nature of CP violation, $m_b$
enters as a crucial input parameter \cite{BSU}.
Thus, precise knowledge of $m_b$ is essential for the interpretation of the
experimental data.
On the other hand, the top-quark mass $m_t$ is one of the key parameters in
the precision tests of the standard model of the electroweak interactions and
in the search for {\it new physics} at a future $e^+e^-$ linear collider.
Furthermore, the study of $t\bar t$ threshold production should even allow us
to probe Higgs-boson-induced effects \cite{StrPes}.
Besides its phenomenological importance, the heavy-quarkonium system is also
very interesting from the theoretical point of view because it possesses a
highly sophisticated multiscale dynamics and its study demands the full power
of the effective-field-theory approach.
Equipped with reliable perturbative results and experimental data on
heavy-quarkonium observables, one can test the effects and structure of the
nonperturbative QCD vacuum.

The binding energy of the bound state and the value of its wave function at
the origin are among the characteristics of the heavy-quarkonium system that
are of primary phenomenological interest.
The former determines the mass of the bound-state resonance, while the latter
controls its production and annihilation rates. 
Recently, the heavy-quarkonium spectrum has been computed through
${\cal O}(\alpha_s^5m_q)$ \cite{KPSS1,PenSte} including the third-order
correction to the Coulomb approximation.
On the other hand, as for the wave function at the origin, a complete result
is so far only available through ${\cal O}(\alpha_s^2)$ \cite{KPP}. 
The ${\cal O}(\alpha_s^2)$ correction has turned out to be so sizeable that
the feasibility of an accurate perturbative analysis was challenged
\cite{gang}, and it appears indispensable to gain full control over the next
order.
Only the double-logarithmic third-order correction, of
${\cal O}(\alpha_s^3\ln^2\alpha_s)$, is available so far \cite{KniPen2}. 
In this Letter, we take the next step and calculate the single-logarithmic
${\cal O}(\alpha_s^3\ln\alpha_s)$ correction.
As a by-product of our analysis, we obtain the three-loop anomalous dimensions
of the nonrelativistic vector and pseudoscalar currents, which constitute
central ingredients for the renormalization-group improvement of the effective
theory of nonrelativistic QCD (NRQCD) \cite{LMR,HMST,Pin2}.
The main results are given by Eqs.~(\ref{cln}), (\ref{ndep}), and
(\ref{gamthr}).
As for the calculation, we follow the general approach of Ref.~\cite{KPSS1}
(see also Ref.~\cite{PinSot2}).
It is based on the nonrelativistic effective-theory concept \cite{CasLep} in
its potential-NRQCD (pNRQCD) incarnation \cite{PinSot1} implemented with the
threshold-expansion technique \cite{BenSmi}.

Let us focus on two examples of paramount phenomenological relevance: the
leptonic decays of the $\Upsilon(1S)$ resonance and the threshold production
of top quark-antiquark pairs in $e^+e^-$ annihilation.
Both processes are essentially photon mediated and thus governed by the
electromagnetic quark current $j_\mu=\bar q\gamma_\mu q$.
Within the effective theory, $j_\mu$ has the following decomposition in terms
of operators constructed from the nonrelativistic quark and antiquark
two-component Pauli spinors $\psi$ and $\chi$ \cite{CasLep}:
\begin{equation}
j_i=c_v(\mu)\psi^\dagger\sigma_i\chi+{d_v(\mu)\over6m_q^2}
\psi^\dagger\sigma_i\mbox{\boldmath$D$}^2\chi
+\ldots,
\label{vcurr}
\end{equation}
where $\mu$ is the renormalization scale, $\mbox{\boldmath$D$}$ are the space
components of the gauge-covariant derivative involving the gluon fields, and
the ellipsis stands for operators of higher mass dimension. 
The Wilson coefficients $c_v(\mu)$ and $d_v(\mu)$ may be evaluated as series
in $\alpha_s(\mu)$ and represent the contributions from the hard modes (where
energy and three-momentum scale like $m_q$) that have been
{\it integrated out}. 
They are computed in full QCD for on-shell on-threshold external (anti)quark
fields and are logarithmic functions of $\mu/m_q$. 
Also integrating out the soft (energy and three-momentum scale like $m_qv$,
where $v$ is the heavy-quark velocity) modes and the potential (energy scales
like $m_qv^2$, while three-momentum scales like $m_qv$) gluons yields the
effective Hamiltonian of pNRQCD, which contains the potential (anti)quarks and
the ultrasoft (energy and three-momentum scale like $m_qv^2$) gluons as active
particles. 
The dynamics of the nonrelativistic potential heavy-quark-antiquark pair in
pNRQCD is governed by the corresponding effective Schr\"odinger equation and
its multipole interactions with the ultrasoft gluons.
The effective-theory expression for the partial decay width of
$\Upsilon(1S)\to l^+l^-$ reads \cite{KPP}
\begin{equation}
\Gamma_1=\Gamma_1^{\rm LO}\rho_1\!
\left[c_v^2(m_b)+{C_F^2\alpha_s^2\over 12}\!c_v(m_b)\left(d_v(m_b)+3\right)
+\ldots\right]\!,
\label{gamll}
\end{equation}
with $\Gamma_1^{\rm LO}=4\pi N_cQ_b^2\alpha^2|\psi_1^C(0)|^2/
\left(3m_b^2\right)$
and $\rho_1=|\psi_1(0)|^2/|\psi_1^C(0)|^2$, where $N_c=3$, $Q_q$ is the
fractional electric charge of quark $q$, $\alpha$ is Sommerfeld's
fine-structu\-re constant, 
$\psi_1(\mbox{\boldmath$x$})$ is the ground-state wave function as
computed in pNRQCD, and $\psi_1^C(\mbox{\boldmath$x$})$ is the Coulomb 
solution, which incorporates the leading binding effects and about which the
perturbative expansion of $\psi_1(\mbox{\boldmath$x$})$ is constructed.
For arbitrary principal quantum number $n$, we have
$\left|\psi^C_n(0)\right|^2=C_F^3\alpha_s^3m_q^3/(8\pi n^3)$, where
$C_F=(N_c^2-1)/(2N_c)$.
Here and in the following, $\alpha_s(\mu)$ is to be evaluated at the soft
normalization scale $\mu_s=C_F\alpha_s(\mu_s)m_q$ whenever its argument is
omitted.
Nonperturbative contributions to Eq.~(\ref{gamll}) are ignored.
The leading one, due to the gluon condensate of the vacuum, may be found in
Ref.~\cite{TitYnd}.
It is quite sizeable and out of control for higher resonances.
A reliable quantitative estimate of the nonperturbative contributions to
Eq.~(\ref{gamll}) can only be obtained through lattice simulations.
On the other hand, to keep the nonperturbative effects under control, one can
employ nonrelativistic $\Upsilon$ sum rules \cite{NSVZ} based on the
global-duality concept.

In the top-quark case, the nonperturbative effects are negligible.
However, the effect of the top-quark total decay width $\Gamma_t$ has to be
properly taken into account \cite{FadKho}, as it is relatively large and
smears out the Coulomb-like resonances below threshold.
The NNLO\footnote{In the effective-theory framework, one has two expansion
parameters, $\alpha_s$ and $v$, and the corrections are classified according
to the total power of $\alpha_s$ and $v$ as leading order (LO),
next-to-leading order (NLO), NNLO, N$^3$LO, {\it etc}.} analysis of the cross
section \cite{gang} shows that only the ground-state pole gives rise to a
prominent resonance.
The value of the normalized cross section  
$R=\sigma(e^+e^-\to t\bar t)/\sigma(e^+e^-\to\mu^+\mu^-)$ at the resonance
energy is dominated by the contribution from the {\it would-be} toponium
ground-state, which is of the form
\begin{equation}
R_1=R_1^{\rm LO}\rho_1\!
\left[c_v^2(m_t)+{C_F^2\alpha_s^2\over 12}\!c_v(m_b)\left(d_v(m_b)+3\right)
\!+\ldots\right]\!,
\label{ree}
\end{equation}
with
$R_1^{\rm LO}=6\pi N_cQ_t^2|\psi_1^C(0)|^2/\left(m_t^2
\Gamma_t\right)$.
The contributions from the higher Coulomb-like poles and the continuum are not
included in Eq.~(\ref{ree}), and we postpone the complete analysis to a future
publication. 
It is understood that $\alpha$ appearing in $\Gamma_1^{\rm LO}$ and
$R_1^{\rm LO}$ is to be evaluated at the mass scale of the respective
resonance.

Starting from ${\cal O}(\alpha_s^2)$, $c_v(\mu)$ is infrared (IR) divergent. 
This divergence arises in the process of scale separation and is canceled
against the ultraviolet (UV) one of the effective-theory result for the wave
function at the origin. 
In our approach, dimensional regularization with $d=4-2\epsilon$ space-time
dimensions is used to handle the divergences, and the formal expressions
derived from the Feynman rules of the effective theory are understood in the
sense of the threshold expansion.
This formulation of effective theory possesses two crucial virtues: the
absence of additional regulator scales and the automatic matching of the
contributions from different scales.
For convenience, we subtract the IR and UV poles in $\epsilon$ according to
the modified minimal-subtraction ($\overline{\rm MS}$) prescription and set
$\mu=m_q$, so that $c_v(m_q)$ is devoid of logarithms.
The latter is known through ${\cal O}(\alpha_s^2)$ and reads \cite{CzaMel1}
\begin{eqnarray}
\lefteqn{c_v(m_q)=1-{\alpha_s(m_q)\over\pi}2C_F
+\left({\alpha_s(m_q)\over\pi}\right)^2\left[\left(-{151\over72}
\right.\right.}
\nonumber\\
&&{}+\left.{89\pi^2\over144}-{5\pi^2\over6}\ln2-{13\over4}\zeta(3)\right)C_AC_F
+\left({23\over8}-{79\pi^2\over36}\right.
\nonumber\\
&&{}+\left.\pi^2\ln2-{1\over2}\zeta(3)\right)C_F^2
+\left({22\over9}-{2\pi^2\over9}\right)C_FT_F
\nonumber\\
&&{}+\left.{11\over18}C_FT_Fn_l\right]+\ldots,
\end{eqnarray}
where $\alpha_s$ is renormalized in the $\overline{\rm MS}$ scheme, $C_A=N_c$,
$T_F=1/2$, $n_l$ is the number of light-quark flavors, and $\zeta(x)$ is
Riemann's $\zeta$ function with value $\zeta(3)=1.202057\ldots$. 
To the order considered, we have $d_v(m_q)=1$.

The corrections to $|\psi_1^C(0)|^2$ read
\begin{eqnarray}
\lefteqn{\rho_1=1+{\alpha_s\over\pi}
\left[\left(4-{2\pi^2\over3}\right)\beta_0+{3\over4}a_1\right]}
\nonumber\\
&&{}+\left({\alpha_s\over\pi}\right)^2\left\{
\left[-C_AC_F+\left(-2+{2\over3}S(S+1)\right)C_F^2\right]\right.
\nonumber\\
&&{}\times\pi^2\ln(C_F\alpha_s)
+\left(-{5\pi^2\over3}+20\zeta(3)+{\pi^4\over9}\right)\beta_0^2
\nonumber\\
&&{}+\left(4-{2\pi^2\over3}\right)\beta_1
+\left({5\over2}-{2\pi^2\over3}\right)\beta_0a_1
+{3\over16}a_1^2+{3\over16}a_2
\nonumber \\
&&{}+\left.{9\pi^2\over4}C_AC_F
+\left({33\pi^2\over8}-{13\pi^2\over9}S(S+1)\right)C_F^2\right\}
\nonumber\\
&&{}+{\alpha_s^3\over\pi}\left\{\left[\left(-2C_AC_F+
\left(-4+{4\over3}S(S+1)\right)C_F^2\right)\beta_0\right.\right.
\nonumber\\
&&{}-\left.{2\over3}C_A^2C_F
+\left(-{41\over12}+{7\over12}S(S+1)\right)C_AC_F^2-{3\over 2}C_F^3\right]
\nonumber\\
&&{}\times\ln^2\left(C_F\alpha_s\right)
+{\cal C}_1\ln\left(C_F\alpha_s\right)+\ldots\bigg\}+\ldots,
\label{wf}
\end{eqnarray}
where $\beta_i$ is the $(i+1)$-loop coefficient of the QCD $\beta$ function
($\beta_0=11C_A/12-T_Fn_l/3,\ldots$) and
$a_i$ parameterizes the $i$-loop correction to the Coulomb potential
($a_1=31C_A/9-20T_Fn_l/9,\ldots$ \cite{Sch}).
For the processes under consideration, the total spin of the quark-antiquark
pair is $S=1$.
Nevertheless, we retain the full $S$ dependence, so that our result is also
applicable to processes with $S=0$, such as the decay $\eta_b\to\gamma\gamma$
or the production process $\gamma\gamma\to t\bar t$ at a future high-energy
photon collider, which is dominated by the $S$ wave \cite{PenPiv2}.
The corrections through ${\cal O}(\alpha_s^2)$ have been derived in
Ref.~\cite{KPP} for arbitrary $n$.
The ${\cal O}(\alpha_s^3\ln^2\alpha_s)$ correction has been obtained in
Ref.~\cite{KniPen2}.
In this Letter, we present the ${\cal O}(\alpha_s^3\ln\alpha_s)$ correction by
specifying the missing coefficient ${\cal C}_1$ in Eq.~(\ref{wf}).

The origin of the logarithmic corrections is the presence of several scales in
the threshold problem.
A logarithmic integral between different scales yields a term proportional to
$\ln v$, which becomes $\ln\alpha_s$ for bound states that are approximately
Coulombic, so that $v\propto\alpha_s$.
In effective-theory calculations, the scale defining the upper (lower) limit
of a logarithmic integral is set to infinity (zero), which induces a UV (IR)
divergence.
Thus, the logarithmic corrections can be identified with the effective-theory
singularities, which dramatically simplifies the calculation.
Our analysis proceeds along the lines of Ref.~\cite{KniPen3} (see also
Ref.~\cite{Hil}), where similar corrections have been considered for the QED
bound-state of positronium.
In the calculation, we employ the N$^3$LO effective Hamiltonian derived in
Ref.~\cite{KPSS1} and take into account the retardation effects due to the
chromoelectric dipole interaction of the heavy-quark-antiquark pair with the
dynamical ultrasoft gluons studied in Refs.~\cite{KPSS1,KniPen1}.
Our result reads
\begin{eqnarray}
\lefteqn{{\cal C}_1=\left[\left(-3+{2\pi^2\over3}\right)C_AC_F+
\left({4\pi^2\over3}-\left({10\over9}+{4\pi^2\over9}
\right)
\right.\right.}
\nonumber\\
&&{}\times
\!S(S+1)\bigg)
C_F^2\bigg]\beta_0+\!\left[-{3\over4}C_AC_F+\left(-{9\over4}
\!+{2\over3}S(S+1)\right)
\right.
\nonumber\\
&&{}
\times C_F^2\bigg]a_1
+{1\over4}C_A^3
+\left({59\over36}-4\ln2\right)C_A^2C_F
+\left({143\over36}\right.
\nonumber \\
&&{}-{4}\ln 2-{19\over108} S(S+1)\bigg)C_AC_F^2
+\left(-{35\over18}+8\ln2-\!{1\over 3}
\right.
\nonumber \\
&&{}
\times\! S(S+1)\bigg)C_F^3
+\!\left(-{32\over15}+\!2\ln2+\left(1-\!\ln2\right)S(S+1)\!\right)
\nonumber\\
&&{}
\times\! C_F^2T_F
+\!{49\over36}C_AC_FT_Fn_l
+\!\left(\!{8\over9}-\!{10\over27}S(S+1)\!\right)\!C_F^2T_Fn_l.
\nonumber\\
&&{}\label{cln}
\end{eqnarray}
For the analysis of $\Upsilon$ sum rules and $t\bar t$ threshold production,
one needs the extension of this result to arbitrary $n$, which, leaving aside
the trivial $n$ dependence of $|\psi_n^C(0)|^2$, is given by
\begin{eqnarray}
\lefteqn{{\cal C}_n={\cal C}_1+
\left[-2C_AC_F+\left(-4+{4\over3}S(S+1)\right)C_F^2\right]\beta_0}
\nonumber\\
&&{}\times\left(\ln n+\Psi_1(n+1)-2n\Psi_2(n)-3+\gamma_E+{\pi^2\over 3}
+{2\over n}\right)
\nonumber\\
&&{}+\left[{4\over3}C_A^2C_F+\left({41\over6}-{7\over6}S(S+1)\right)C_AC_F^2
+3C_F^3\right]
\bigg(\ln n
\nonumber\\
&&{}-\left.\Psi_1(n)-1-\gamma_E+{1\over n}\right)
+\left({5\over3}-{5\over3n^2}\right)C_AC_F^2,
\label{ndep}
\end{eqnarray}
where $\Psi_n(x)=d^n\ln\Gamma(x)/dx^n$, 
$\Gamma(x)$ is Euler's $\Gamma$ function,
and $\gamma_E=0.577216\ldots$ is Euler's constant.

The structure of the IR singularities in the Wilson coefficients can be read
off from the UV-singular part of the effective-theory result for the wave
function at the origin.  
In this way, we find the $\mu$ dependence of $c_v(\mu)$ to be
\begin{eqnarray}
c_v^2(\mu)&=&c_v^2(m_q)\left\{1+\alpha^2_s(\mu)\gamma_v^{(2)}
\ln{\mu^2\over m_q^2}+{\alpha_s^3(\mu)\over\pi}\right.
\nonumber\\
&&{}\times\left[\left(-{3\over2}\beta_0\gamma_v^{(2)}+\gamma_v^{(3)}\right)
\ln^2{\mu^2\over m_q^2}
+\bigg(-2\beta_0\gamma_v^{(2)}\right.
\nonumber\\
&&{}+\left.\left.\left.{4\over3}\gamma_v^{(3)}
+\gamma_v^{\prime(3)}\right)
\ln{\mu^2\over m_q^2}+\ldots\right]+\ldots\right\},
\end{eqnarray}
where the two- \cite{CzaMel1} and three-loop anomalous dimensions of the
nonrelativistic vector current  are
\begin{eqnarray}
\lefteqn{\gamma_v^{(2)}
=-{1\over2}C_AC_F+\left(-1+{1\over3}S(S+1)\right)C_F^2,}
\label{gamtwo}\\
&&\gamma_v^{(3)}=
-{1\over8}C_A^2C_F+\!\left(\!-{29\over32}+\!{7\over32}S(S+1)\!\right)C_AC_F^2
-\!{5\over16}C_F^3,
\nonumber\\
&&\gamma_v^{\prime(3)}=
\left[-{1\over 2}C_AC_F
+\left(2-{11\over 9}S(S+1)
\right)C_F^2\right]\beta_0
\nonumber \\
&&{}
+\left(-{3\over 8}
+{1\over 12}S(S+1)\right)C_F^2a_1-\left({53\over 72}
+{3\over 2}\ln 2\right)C_A^2C_F
\nonumber \\
&&{}
+\left(
-{107\over 36}-{3\over 2}\ln 2+{373\over 432}S(S+1)\right) C_AC_F^2
+\left(-{7\over 8}\right.
\nonumber \\
&&{}
\left.+3\ln 2 -{1\over 4}S(S+1)\right)C_F^3+\!\left(-{8\over 5}
+{3\over 2}\ln 2
\! +\left(\!{3\over 4}-\!{3\over 4}\ln 2\!\right)
\right.
\nonumber \\
&&{}
\times S(S+1)\bigg)C_F^2T_F+{49\over 72}C_AC_FT_Fn_l+
\left({4\over 9}-{5\over 27}\right.
\nonumber \\
&&{}
\times S(S+1)\bigg)
C_F^2T_Fn_l,
\label{gamthr}
\end{eqnarray}
with $S=1$.
We retain the full $S$ dependence in Eqs.~(\ref{gamtwo}) and (\ref{gamthr})
because, for $S=0$, they give the anomalous dimension of the nonrelativistic
pseudoscalar current $\psi^\dagger\chi$, which is relevant for two-photon
processes.
Note that the $S=0$ result for $C_1$ and  $\gamma_v'^{(3)}$
corresponds to the  $d$-dimensional spinor algebra
in a regularization scheme adopted in \cite{CzaMel}
for the calculation of the  two-loop hard
corrections to the two-photon  heavy quarkonium production
and annihilation.

Let us now explore the numerical significance of our results.
Putting everything together and substituting constants by their approximate
numerical values, Eq.~(\ref{gamll}) becomes
\begin{eqnarray}
\Gamma_1&\approx&\Gamma_1^{\rm LO}
\left(1-1.70\,\alpha_s(m_b)-7.98\,\alpha_s^2(m_b)+\ldots\right)
\nonumber\\
&&{}\times\left(1-0.30\,\alpha_s-5.19\,\alpha_s^2\ln\alpha_s+17.16\,\alpha_s^2
\right.
\nonumber\\
&&{}-\left.14.38\,\alpha_s^3\ln^2\alpha_s-0.165\,\alpha_s^3\ln\alpha_s
+\ldots\right).
\label{gamser}
\end{eqnarray}
Note that the NNLO contribution due to the second term in the square brackets
of Eq.~(\ref{gamll}) is included in the third factor on the right-hand side of
Eq.~(\ref{gamser}), as the corresponding appearance of $\alpha_s$ is of
nonrelativistic origin and normalized at the soft scale.
Evaluating this using $\alpha_s(M_Z)=0.1185$ and $m_b=5.3$~GeV \cite{PenSte},
we obtain
\begin{equation}
\Gamma_1\approx\Gamma^{\rm LO}_1(1-\!0.449_{\rm NLO}
+\!1.771_{\rm NNLO}-\!0.766_{\rm N^3LO'}+\!\ldots).
\label{gamnum}
\end{equation}
where only the logarithmic N$^3$LO terms are retained, which is
indicated by the prime on the subscript.
A similar analysis of Eq.~(\ref{ree}) with $m_t=174.3$~GeV yields 
\begin{equation}
R_1\approx R_1^{\rm LO}(1 -\!0.244_{\rm NLO}
+\!0.438_{\rm NNLO}-\!0.196_{\rm N^3LO'}\!+\!\ldots).
\label{rnum}
\end{equation}
Without the ${\cal O}(\alpha_s^3\ln\alpha_s)$ term, the N$^3$LO contributions
in Eqs.~(\ref{gamnum}) and (\ref{rnum}) read $-0.560$ and $-0.148$,
respectively.
We learn the following:
(i) while the coefficients in Eq.~(\ref{gamser}) and the analogous series for
$R_1$ sharply increase in magnitude as we pass from NLO to NNLO, this
disquieting trend discontinues as we move on to N$^3$LO;
(ii) the coefficients of the known NNLO and N$^3$LO terms are typically of
order 10.
These observations suggest that the magnitude of the coefficient of the
missing non-logarithmic ${\cal O}(\alpha_s^3)$ term is unlikely to exceed this
characteristic benchmark by far.
This term would then be expected to yield corrections of order 25\% and 3\% to
$\Gamma_1$ and $R_1$, respectively.
This provides us with an estimate of the residual uncertainty of our
approximation as far as the perturbative QCD corrections are concerned.
Moreover, the absence of a rapid growth of the coefficients along with the
alternating-sign character of the series suggest that the higher-order
corrections are likely to be below these estimates.
These observations can be substantiated by investigating the scale dependence
of $\Gamma_1$ and $R_1$.
In fact, the shifts in these quantities due to a variation of $\mu_s$ by a 
factor of 2 are reduced from 50\% and 13\% to 19\% and 9\%, respectively, as 
we pass from NNLO to N$^3$LO.
The latter values are in the same ballpark as the theoretical uncertainties
estimated above.
This renders a reliable perturbative description of heavy-quarkonium
production and annihilation feasible.

To conclude, we computed the logarithmically enhanced N$^3$LO corrections to
the heavy-quarkonium production and annihilation rates. 
Our results provide a useful hint on the general structure of the high-order
corrections and open a new perspective for the theoretical analysis.
Together with the ${\cal O}(\alpha_s^5m_q)$ result for the heavy-quarkonium
spectrum \cite{KPSS1,PenSte}, they constitute central ingredients for the
high-precision analysis of $\Upsilon$ sum rules and $t\bar t$ threshold
production in $e^+e^-$ and $\gamma\gamma$ scattering.
Calculation of the remaining non-logarithmic N$^3$LO term appears to be
mandatory for reducing the theoretical uncertainty further.
Another challenging problem is to complete the resummation of the
next-to-next-to-leading logarithms \cite{HMST}.
Our derivation of the three-loop anomalous dimension of the nonrelativistic
vector current marks a major step in this direction.

A.A.P. acknowledges discussions with A. Hoang. 
This work was supported in part by DFG Grant No.\ KN~365/1-1 and BMBF Grant
No.\ 05~HT1GUA/4.
The work of V.A.S. was supported in part by RFBR Project No.\ 01-02-16171,
Volkswagen Foundation Contract No.\ I/77788, and INTAS Grant No.\ 00-00313.

\vspace{5mm}
\noindent
{\large \bf Note added:}

\noindent
In a recent paper \cite{Hoang} the infrared structure
of the three-loop corrections to the heavy quark pair
production current was analyzed in  the effective theory framework. 
On the basis  of this investigation the next-to-next-to leading logarithmic 
corrections to the heavy quarkonium threshold production
were partially resummed. By expanding  this result in $\alpha_s$
the ${\cal O}(\alpha_s^3\ln\alpha_s)$ correction  
to the heavy-quarkonium production and annihilation rates
can be obtained. The result announced in \cite{Hoang}
for the $S=1$ spin-one process agrees with  our result for $C_1$, 
Eq.~(\ref{cln}).

\end{document}